\documentstyle[twoside,fleqn,espcrc2,epsf]{article}

\newcommand{\AmS}{{\protect\the\textfont2
  A\kern-.1667em\lower.5ex\hbox{M}\kern-.125emS}}

\begin{document}
\title{
\vspace*{-1.5cm}
\begin{flushright}
{\small
KANAZAWA 98-14, 
HUB-EP 98/65}
\end{flushright}
\vspace*{0.5cm}
 Abelian monopoles in finite temperature lattice gauge fields:\\  
 Classically perfect action, smoothing and various Abelian gauges 
}

\author{
E.-M. Ilgenfritz\address{Institute for Theoretical Physics, 
University of Kanazawa, Japan},
H. Markum\address{Institut f\"ur Kernphysik, TU-Wien, Austria}, 
M. M\"uller-Preussker\address{Institut f\"ur Physik, 
Humboldt-Universit\"at zu Berlin, Germany}  
and S. Thurner$^{\rm b,c}$ 
}

\begin{abstract}
Using the renormalization group motivated smoothing technique, 
the large scale structure of lattice
configurations at finite temperature is characterized in terms 
of Abelian monopoles identified in the 
maximally 
Abelian, the Laplacian Abelian, and the Polyakov gauge. 
Abundance and anisotropy of monopoles at deconfinement and gauge 
invariant properties like local non-Abelian action
and topological density are studied. 
Monopoles are predominantly found in regions of large 
action and topological charge, rather independent of the chosen gauge. 
\end{abstract}

\maketitle

Confinement in non-Abelian gauge theories has a popular 
explanation within
the dual superconductor picture. In this scenario 
the condensation of Abelian monopoles leads to confinement 
of color charges through a dual Meissner effect. 
The mechanism 
has been substantiated  by a large number of lattice studies. 
Abelian monopoles 
in the confinement phase, 
obtained from Abelian projection in an appropriate gauge and
representing links on the dual lattice, have been shown to
percolate through the $4D$ volume and 
to be responsible for a dominant contribution to the string 
tension \cite{SUZU}. 
\begin{figure}[h] 
\vspace{-0.6cm}
\begin{tabular}{c}
\epsfxsize=7.5cm\epsfysize=7.5cm\epsffile{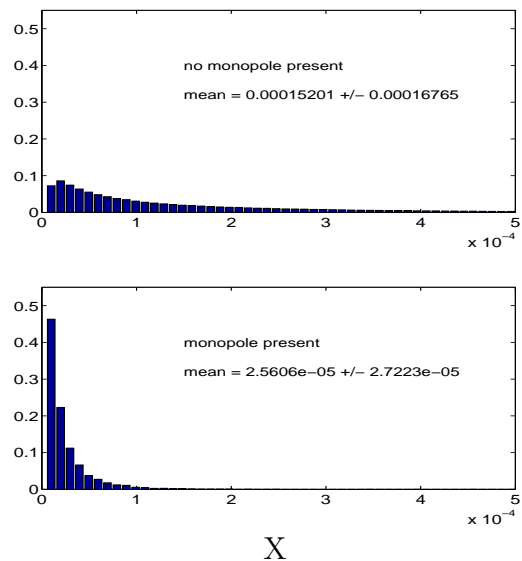}\\
{\large X}\\
\end{tabular}
\vspace{-1.0cm}
\caption{Probability distribution of the local norm $X=||\Phi_x||$ of the
	 auxiliary Higgs field of LAG, in 
         the case of absence (top) or presence (bottom) of 
         a DGT monopole. 
}
\label{fig1}
\vspace{-1cm}
\end{figure}

\begin{figure}[] 
\epsfxsize=7.0cm\epsffile{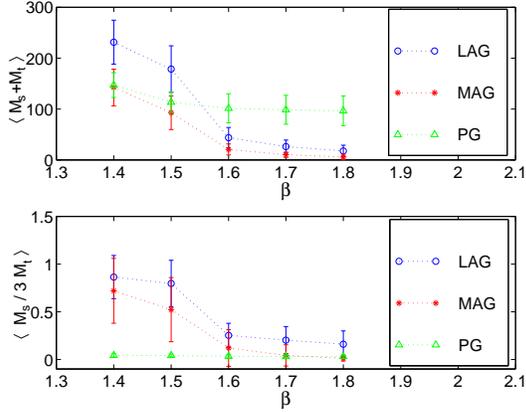}
\vspace{-1.1cm}
\caption{Total monopole length (top) and space-time asymmetry (bottom) 
         as a function of 
         $\beta$ ($\beta_c = 1.545(10)$) 
         for  monopoles in different gauges.}
\label{fig2}
\vspace{-0.7cm}
\end{figure}
Studying 
creation operators of monopoles, evidence 
for their condensation was found, independently of the 
gauge chosen \cite{DIGI2}. 
On the other hand, 
lengths and locations of monopole  trajectories 
{\it do} depend on the selected gauge. 
In order to point out other potential differences between monopoles
corresponding to various gauges, we concentrate here on aspects 
of temperature dependence at the phase transition and of 
gauge invariant characteristics 
such as action and topological charge. To be able to do this we 
have studied 
the semiclassical vacuum structure.

\begin{figure}[htb] 
\begin{tabular}{c}
\hspace{-0.3cm} \epsfxsize=7.4cm\epsffile{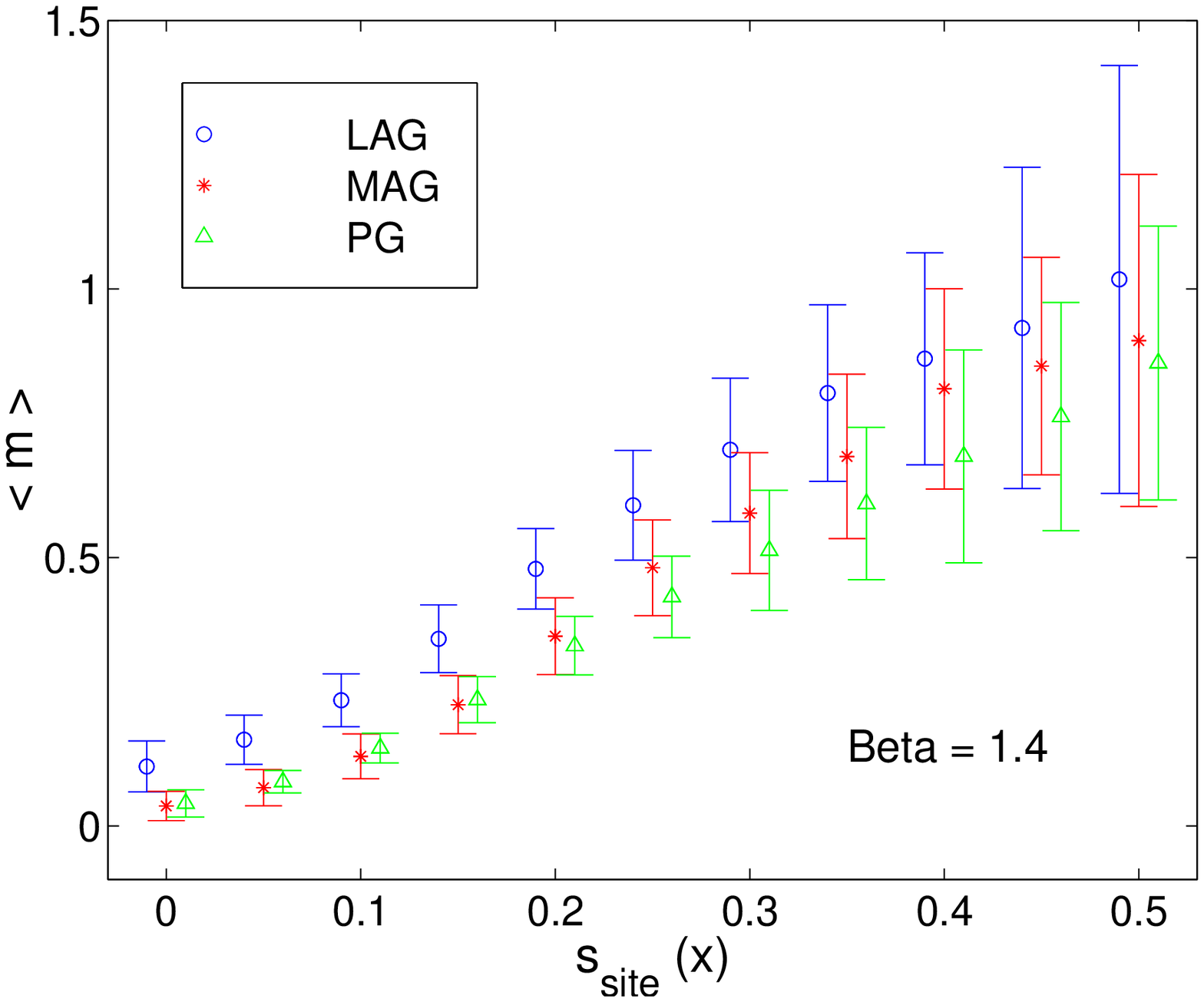}\\
\hspace{-0.7cm} \epsfxsize=7.0cm\epsffile{  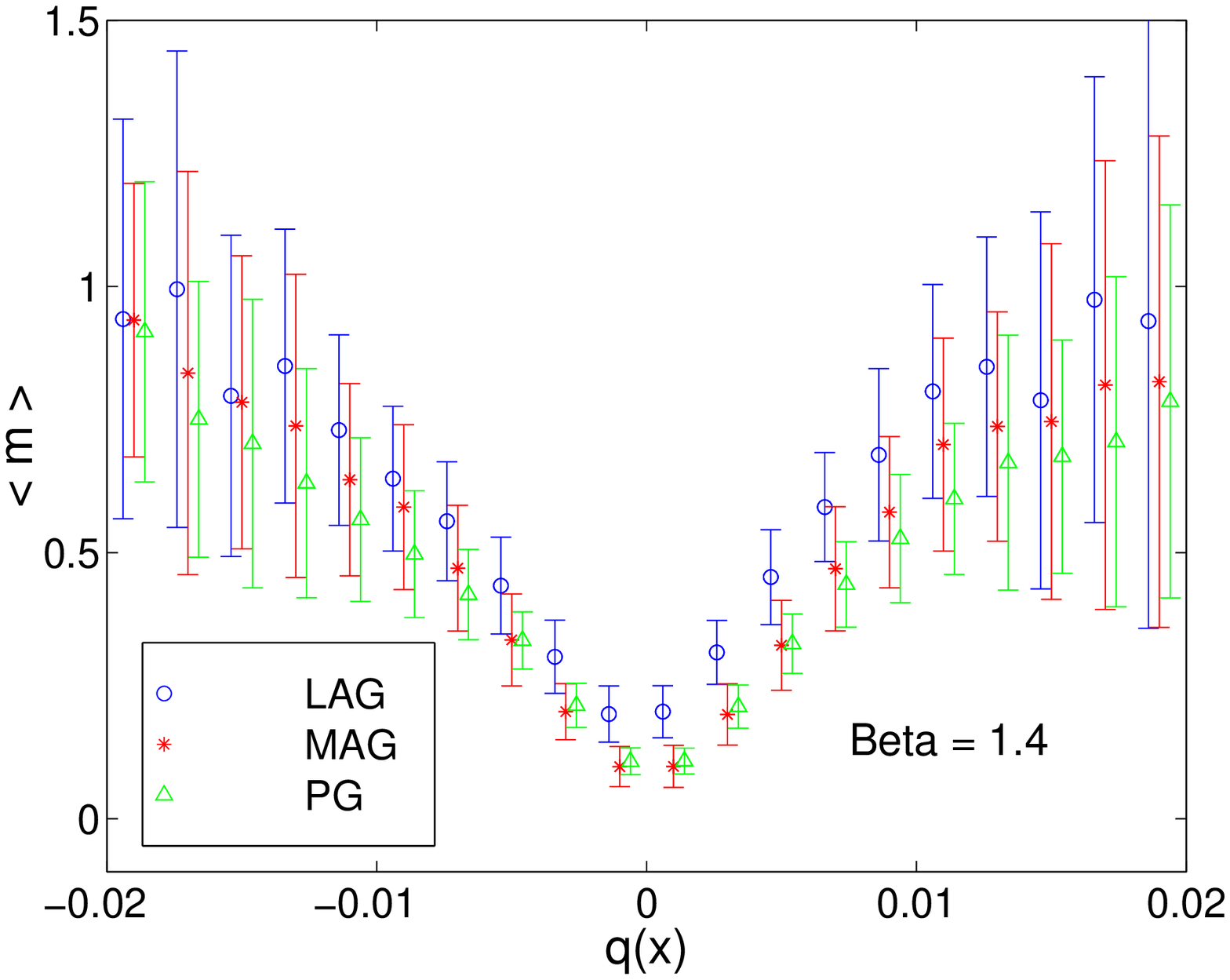}\\
\end{tabular}
\vspace{-1.0cm}
\caption{Average occupation number of monopoles $<m>$ at sites 
with action density
$s_{\rm site}$ (top) and topological charge density $q$ 
(bottom) in the confinement phase.}
\label{fig3}
\vspace{-0.9cm}
\end{figure}
To resolve the semiclassical structure of gauge fields, 
mostly the `cooling' method has been used. However, even improved versions 
of this method loose small instantons and instanton-antiinstanton
pairs, and will destroy monopole percolation as well.
To overcome these problems we employ a method of constrained 
smoothing \cite{FEUE} which is based on the concept of perfect actions 
\cite{HASE}. This method does not drive configurations into classical 
fields but keeps large scale structures as they are 
deformed by quantum interactions. 
There is a well-defined scale above which the
semiclassical structure of the raw configuration is preserved\footnote{The 
iterative application of this method, `cycling' \cite{BOUL},
obscures the very idea of a definite blocking scale while it is
still preserving long range physics rather well.}.

We use a simplified fixed-point action  \cite{PRD98} for Monte Carlo and
for `constrained smoothing' before a configuration is analyzed.  
Simulations were done on a $12^3\times 4$ lattice. 
Observables were computed on 
50 independent configurations, for each of the $\beta$ values considered. 
The Abelian gauges considered here are the 
maximally
Abelian (MAG) \cite{KRON87}, the Laplacian 
Abelian (LAG) \cite{SIJS} and the Polyakov gauge (PG).
The MAG can be considered as the minimization of 
\begin{equation}
F \left[ \Omega \right] =\sum_{x,\mu} \frac{1}{2}\mathrm{Tr} 
\left(\Phi_x - U_{x,\mu} \Phi_{x+\hat\mu} U_{x,\mu}^{\dag}\right)^2  \,\, ,
\end{equation}
with the constraint $||\Phi_x||=1$ where the gauge transformation   
$\Omega_x$ is encoded in $\Phi_x=\Omega_x^{\dag} \sigma_3 \Omega_x$. 
The constraint on $\Phi$ is relaxed in the LAG, so that Eq. (1) 
can be interpreted as the kinetic term of an auxiliary adjoint Higgs field.
Thus gauge fixing reduces to a lowest-eigenvalue problem  
for the covariant lattice Laplacian. If there is no degeneracy,
LAG is globally unambiguous. As in MAG, fixing to LAG means diagonalizing 
$\Phi_x$ which is well-defined if $||\Phi_x||\ne 0$.
(Similarly, PG is enforced by diagonalizing Polyakov lines.) 
DeGrand-Toussaint (DGT) monopoles are then defined by Abelian projection 
from LAG. 
We show in Fig. 1 that the zeros of the Higgs field
corresponding to the gauge fixing singularities are correlated to the 
DGT monopoles where the Higgs modulus is strongly cut off.

Global properties like 
the total monopole loop length and the space-time asymmetry 
are compared in Fig. 2. 
Monopoles from MAG and LAG 
show a similar behaviour across the deconfinement phase
transition.
On the contrary, for the PG monopoles no change is seen in the 
neighborhood of the transition reflecting the 
fact that PG monopoles are static in both phases. 

In Fig. 3 we show the occupation 
probability of a monopole as a function of the local 
action $s_{site}(x)$ and charge $q(x)$ surrounding it, for the
different gauges. This demonstrates 
that the probability of finding monopoles increases 
with the local value of action density/modulus of charge density.
This result is practically independent of the  gauge being used to construct  
the DGT monopoles.

We define an excess action of monopoles by  
\begin{equation}
S_{\rm ex}= 
\frac{< S_{\rm monopole}-S_{\rm no monopole} > }{<S_{\rm no monopole} >} \,\, ,
\end{equation}
where $S_{\rm monopole}$ is the action localized in a three-dimensional 
cube which corresponds to a dual link occupied by a monopole.  
Replacing the action by the modulus of topological charge 
according to L\"uscher's definition we obtain the charge excess $q_{\rm ex}$. 
For details of the definition of the local operators see \cite{PRD98}.
\begin{figure}[htb]
\begin{tabular}{c}
\epsfxsize=7.0cm\epsffile{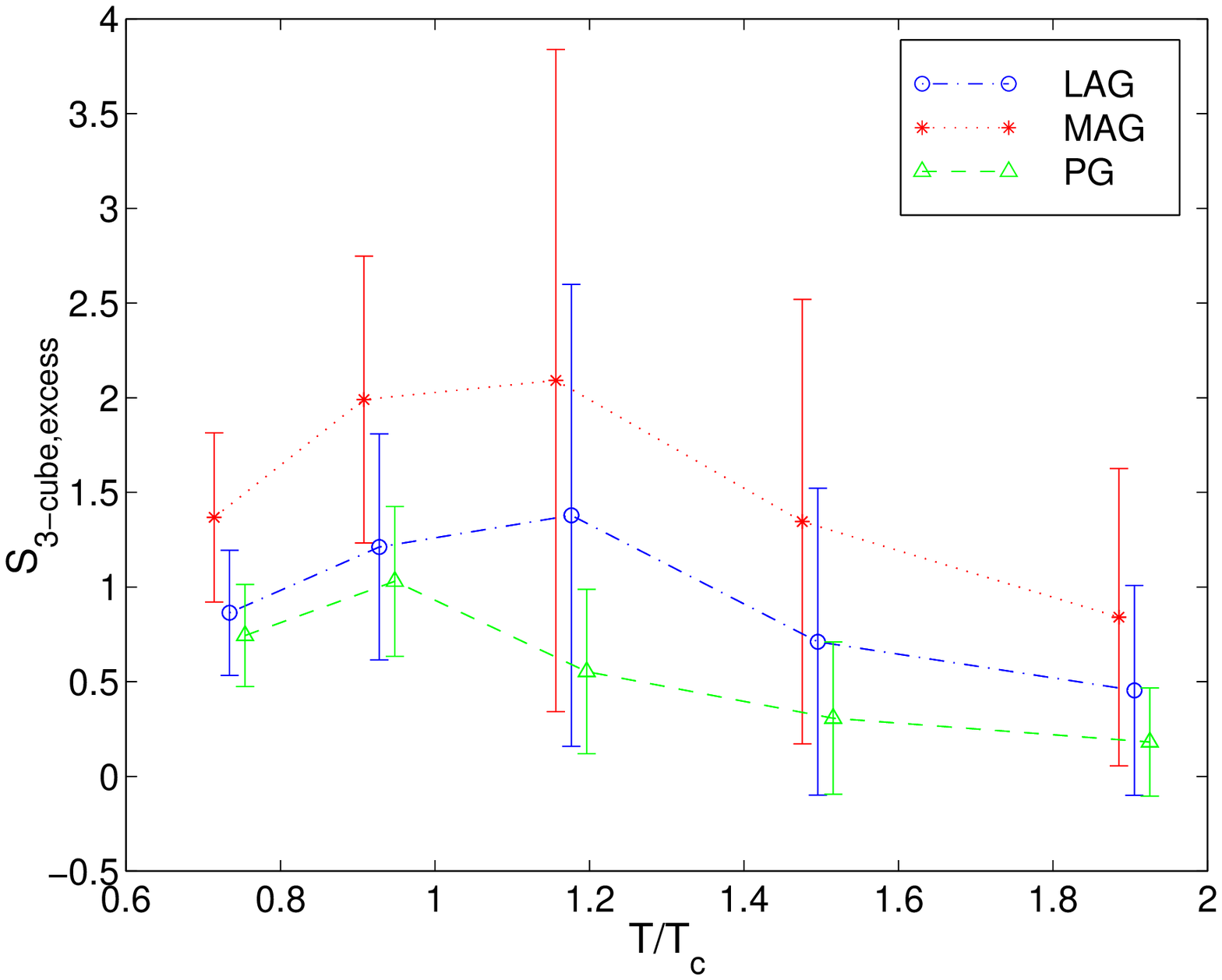}\\
\epsfxsize=7.0cm\epsffile{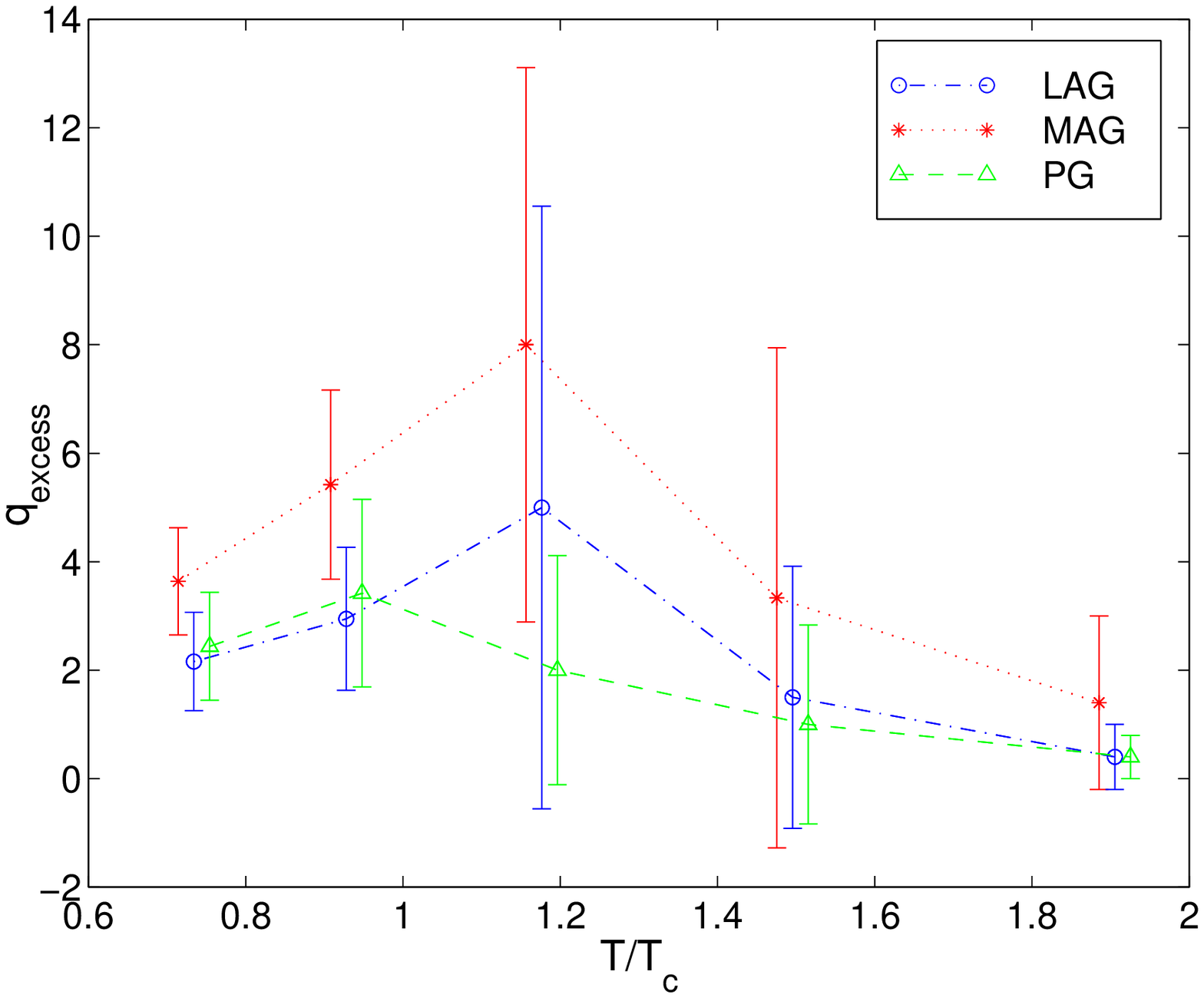}\\

\vspace{-1.5cm}
\end{tabular}
\caption{Excess action (top) and charge (bottom)  as a  function of 
temperature.   }
\label{fig4}

\vspace{-0.8cm}
\end{figure}
Somewhat below
$T_c$ the excess action and charge for the MAG and 
LAG monopoles are above one,
indicating an excess of action of more than a factor of  2 above background. 
The large error bars above $T_c$ reflect the fact that the topological 
activity decreases in the deconfinement phase. Our results 
for the action excess in the confinement phase qualitatively agree
with a recent study without smoothing~\cite{BAKK98}.

Summarizing, we have provided evidence that Abelian monopoles are mainly 
localized in regions, 
which are characterized by enhanced action and topological charge. 
Therefore, monopoles  can, at least to a certain degree, 
be viewed in a gauge invariant language 
as physical objects carrying action and topological charge.

\end{document}